\documentclass[twocolumn,aps,preprintnumbers,amsmath,amssymb,superscriptaddress,nobibnotes]{revtex4}
\usepackage{graphicx}
\usepackage{dcolumn}% Align table columns on decimal point
\usepackage{bm}% bold math
\usepackage{textcomp}
\usepackage{units}
\usepackage{graphics}
\usepackage{longtable}
\begin{document}

\title{Effective beam separation schemes for the measurement of the electric Aharonov-Bohm effect in an ion interferometer}

\author{G. Sch\"{u}tz, A. Rembold, A. Pooch, H. Prochel and A. Stibor}
\email{alexander.stibor@uni-tuebingen.de}
\address{Physikalisches Institut and Center for Collective Quantum Phenomena in LISA$^+$,
Universit\"{a}t T\"{u}bingen, Auf der Morgenstelle 15, 72076 T\"{u}bingen, Germany}

\begin{abstract}
We propose an experiment for the first proof of the type I electric Aharonov-Bohm effect in an ion interferometer for hydrogen. The performances of three different beam separation schemes are simulated and compared. The coherent ion beam is generated by a single atom tip (SAT) source and separated by either two biprisms with a quadrupole lens, two biprisms with an einzel-lens or three biprisms. The beam path separation is necessary to introduce two metal tubes that can be pulsed with different electric potentials. The high time resolution of a delay line detector allows to work with a continuous ion beam and circumvents the pulsed beam operation as originally suggested by Aharonov and Bohm. We demonstrate, that the higher mass and therefore lower velocity of ions compared to electrons combined with the high expected SAT ion emission puts the direct proof of this quantum effect for the first time into reach of current technical possibilities. Thereby a high detection rate of coherent ions is crucial to avoid long integration times that allow the influence of dephasing noise from the environment. We can determine the period of the expected matter wave interference pattern and the signal on the detector by determining the superposition angle of the coherent partial beams. Our simulations were tested with an electron interferometer setup and agree with the experimental results. We determine the separation scheme with three biprisms to be most efficient and predict a total signal acquisition time of only \unit[80]{s} to measure a phase shift from 0 to 2$\pi$ due to the electric Aharonov-Bohm effect. 

\end{abstract}

\maketitle

\section{Introduction}

In the paper of Aharonov and Bohm \cite{Aharonov1959}, two experiments are proposed to prove that the influence of vector and scalar potentials have a direct physical effect on the phase of charged particles, even in absence of any field. Their predictions are known as the magnetic and the electric Aharonov-Bohm effects. Alternatively, they are nominated after Ehrenberg and Siday who were the first revealing this phenomenon \cite{Ehrenberg1949}. 
Soon after, the phase shift resulting from the magnetic Aharonov-Bohm effect could be demonstrated in an electron biprism interferometer \cite{Chambers1960,Mollenstedt1962}. The experiment showed impressively that potentials are not only mathematical constructs to calculate fields, as it was widely believed at that time, but seem to be more fundamental. It encouraged a new community in quantum physics to perform principal tests of the effect, e.g.~by excluding stray fields by a toroidal magnet with a superconducting cladding \cite{Tonomura1986}. In the course of time the field was divided into type I and type II experiments \cite{Batelaan2009}. The original proposed effects are of type I, which means that the electron encounters no magnetic or electric field while it traverses the magnetic vector potential or the electric scalar potential. Type II effects include experiments with electrons \cite{Matteucci1985} or neutrons \cite{Zeilinger1985}, like the Aharonov-Casher \cite{Aharonov1984,Cimmino1989} or the neutron-scalar Aharonov-Bohm effect \cite{Allman1992,Lee1998}. They allow the particle to traverse through a non-vanishing magnetic or electric field, if they do not deflect or delay the wave packet \cite{Batelaan2009}. Until today, only the magnetic Aharonov-Bohm effect could be proven in type~I experiments \cite{Chambers1960,Mollenstedt1962,Tonomura1986}. It was technically not possible to perform a proof of the  type I electric Aharonov-Bohm effect. There are two major reasons. First, electrons emitted from conventional sources are too fast to switch an electric potential on and off quickly enough. And second, implementing a fast pulsing electron source in an interferometer is technically demanding, even if ultrafast pulsed electron sources are available for some years~\cite{Hommelhoff2006}.\\

\begin{figure*}[t]
%\centerline{\scalebox{0.5}{\includegraphics{figure1.eps}}}
\centerline{\scalebox{1.0}{\includegraphics{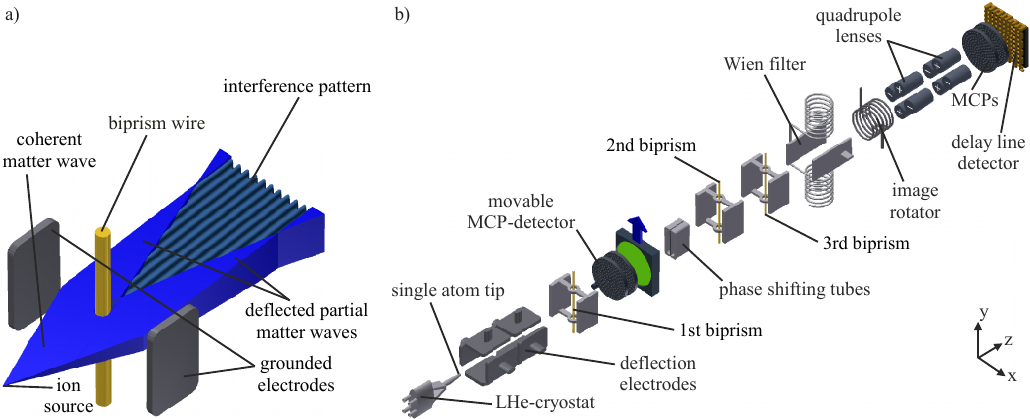}}}
\caption{(Color online) a) Basic scheme for biprism ion or electron interferometry \cite{Mollenstedt1956a}. An electrostatically charged biprism wire separates and recombines a coherent particle wave. b) Setup of the proposed ion interferometer for the measurement of the electric Aharonov-Bohm effect (not to scale) with a beam path separation scheme using three biprisms. Several components are based on the first ion interferometer by Hasselbach et al.~\cite{Hasselbach1998a,Hasselbach1996,Maier1997,Hasselbach2010a} combined with an novel SAT source \cite{Kuo2006a,Kuo2008,Schuetz2014}, the beam path separation, the phase shifting tubes and the time resolving detector.} 
\label{fig:biprism}
\end{figure*}

In this work we propose an experimental scheme to overcome these problems and demonstrate that the first type I measurement of the electric Aharonov-Bohm effect is in direct reach of current technical capabilities. Instead of electrons, as proposed by Aharonov-Bohm, we suggest to use hydrogen ions as particles. Our setup is influenced by the first biprism ion interferometer realized by Hasselbach et al.~\cite{Hasselbach1998a,Hasselbach1996,Maier1997,Hasselbach2010a} and the first electron interferometer with a single atom tip (SAT) beam source \cite{Schuetz2014,Kuo2006a,Kuo2008}. The proposed Aharonov-Bohm experiment implements the intense SAT ion source \cite{Fu2001,Kuo2006a}, a \unit[100]{\textmu m} beam separation scheme and a detector with a high time resolution. The separation is necessary to implement the phase shifting tubes \cite{Aharonov1959,Schmid1985}. The hydrogen ion energy can be controlled between \unit[3.3 and 4.2]{keV} \cite{Kuo2008} and the low energy spread of the SAT is supposed to be in the same order of magnitude as for other gas field ion sources ($<$\unit[1]{eV}) \cite{Kuo2008}. The much heavier and therefore slower ions in combination with modern puls-generators solve the problem of fast switching on the phase shifting tubes in the electric Aharonov-Bohm setup \cite{Aharonov1959}. The high time resolution of modern delay line detectors \cite{Jagutzki2002} of less than \unit[1]{ns} towards a reference signal allows to work with a continuous ion beam and a pulsed detection mode.\\
We will provide detailed classical simulations of the ion beam paths for three different beam separation schemes: two biprisms with a quadrupole, two biprisms with an einzel-lens \cite{Fickler2009} and three biprisms. The simulations allow to determine the superposition angle of the coherent partial beams. The periodicity of the quantum mechanical interference pattern depends on this angle and the ion de Broglie wavelength. Our simulations additionally determine the magnification of the interferogram by quadrupole lenses and the expected count rates on the detector for the three different schemes. A large detection rate of coherent ions is important to avoid long signal integration and the associated dephasing due to disturbances from the environment (temperature drifts, mechanical- and electromagnetic noise \cite{Rembold2014}). Our simulation reveals that the beam path separation scheme with three biprism yields the highest signal rate on the detector. To verify our method we additionally simulated the setup of an electron biprism interferometer \cite{Schuetz2014} and compared our results to the experimental data. 

\section{Proposed experimental setup}
\label{expsetup}

The proposed experimental scheme uses a biprism interferometer to coherently separate and combine charged particle matter waves. The basic principle is shown in fig.~\ref{fig:biprism} a) and was first realized for electrons by M\"ollenstedt et al.~\cite{Mollenstedt1956a}. Thereby an electrostatically charged fine wire, the biprism, is located between two grounded electrodes. It is illuminated coherently by an electron or ion source with a small source size and a low energy spread of the emitted particles. The biprism acts as a beamsplitter with tuneable control of the angle between the partial waves depending on the potential applied on the wire. Therefore, a positive (for electrons) or negative (for ions) potential results in a separation of the matter wave in front of the wire and a recombination afterwards. The split beam paths overlap and interfere with each other. Such a biprism scheme was applied e.g.~by Hasselbach et al.~for the realization of the first ion interferometer for He$^+$-ions \cite{Hasselbach1998a,Hasselbach1996,Maier1997,Hasselbach2010a}. \\
\\
The proposed setup enhances this scheme to enable the first direct measurement of the electric Aharonov-Bohm effect. It includes several additional components, namely a SAT field ion emitter, a second biprism, electrostatic phase shifting tubes, a focusing element, that is either a quadrupole lens, an einzel-lens or a third biprism, and a time resolving delay line detector. The configuration with the three biprisms is shown in fig.~\ref{fig:biprism} b). In the following we provide only a brief description of the components in the interferometer. They are described in detail elsewhere \cite{Schuetz2014,Hasselbach1998a,Hasselbach1996,Maier1997,Hasselbach2010a}.\\
As illustrated in fig.~\ref{fig:biprism} b), the coherent hydrogen ion beam is generated by field ionization of hydrogen gas on a single atom tip \cite{Fu2001,Kuo2006a,Schuetz2014} at which a voltage of \unit[3.8]{kV} is applied. It is cooled by a liquid helium cryostat to $\sim$~\unit[20]{K} and surrounded by hydrogen gas with a pressure between $10^{-6}$ and \unit[$10^{-4}$]{mbar} whereas the amount of ions in the beam increases at higher pressures. However, the application of high pressures above $\sim$ \unit[$10^{-6}$]{mbar} requires the implementation of a differential pumping stage. This would decrease the pressure near the multichannel plate (MCP) detector that could otherwise be harmed and also decrease collisions of the beam with background gas that lower the signal to noise ratio. Referring to the results by Kuo et al~\cite{Kuo2008} there are indications that the source size of the single atom emitter does not vary up to this range of pressure. \\
\\
A MCP detector behind the first biprism is used to control the beam shape prior to the measurement. It can be moved out of the beam path. The beam is adjusted by electrostatic deflector electrodes towards the biprism beam splitter. It consists of a fine, gold-palladium coated, glass fibre \cite{Schuetz2014} on a positive potential. It is set between two grounded plates and divides the ion beam up to \unit[100]{\textmu m} to fit it through two separated tubes. The phase between the two partial waves is shifted by different potentials set on the tubes according to the electric Aharonov-Bohm effect \cite{Aharonov1959}. This process will be explained in detail in section \ref{elABeffect}. In our simulations we apply two rectangular tubes, each with an inner profile of \unit[1.1]{mm} in the $y$-direction and \unit[0.2]{mm} in the $x$-direction. The length in the $z$-direction is \unit[3]{mm}. The tubes wall thickness in the $x$-direction close to the optical axis is very thin, only \unit[2]{\textmu m}, due to the mentioned limited beam path separation. The tubes are arranged left and right of the center at $x = 0$ with an interspace of \unit[2]{\textmu m} for electrical isolation. In our simulation it was filled with vacuum. However, an experimental realization could apply a thin lacquer coat between two metal foils instead to isolate the tubes from each other. Such a configuration was chosen by Schmid et al.~\cite{Schmid1985} where one of the separated electron beam paths was guided through a comparable metal tube. There are different methods to combine the beam paths. Schmid \cite{Schmid1985} used a focusing einzel-lens in combination with a second biprism for the largest coherent electron beam splitting in a biprism interferometer of \unit[300]{\textmu m}. However, his electron interferometer was significantly longer than our proposed setup and therefore more susceptible to mechanical vibrations. Other methods to focus the beam are by a quadrupole lens \cite{Sonnentag2007a} or a second biprism \cite{Hasselbach1993b}, in combination with an additional biprism. We will compare these schemes in simulations presented in section \ref{simulation}.\\ 
\\
Behind these elements the beam traverses a Wien filter that corrects possible longitudinal phase shifts of the ions due to the beam adjustment by the electrostatic deflectors. The Wien filter is applied in most biprism interferometers and explained in detail elsewhere \cite{Nicklaus1993a}. After the superposition of the coherent parts of the divided beams an interference pattern is formed. It is adjusted by an image rotating magnetic coil towards the axis of two magnifying quadrupole lenses. They consist of four electrodes with a length of \unit[20]{mm} for the first quadrupole and \unit[10]{mm} for the second. The pattern is detected by a delay line detector \cite{Jagutzki2002}. Unlike detectors with fluorescent screens in former biprism setups, it has also a temporal resolution besides the spatial one. Such delay line detectors consist of two MCP's and a delay line anode and are commercially available with a temporal resolution of \unit[1]{ns} towards a reference signal. As it will be pointed out in section \ref{elABeffect}, the ability to correlate the incoming ions in the time domain to an external reference pulse is essential for the measurement of the electric Aharonov-Bohm effect in our proposed setup.\\
\\
Matter wave interferometry experiments with large beam path separation require a high degree of stability concerning vibrations and alternating magnetic fields \cite{Rembold2014}. The setup design should therefore consider the electron interferometer configuration realized by Hasselbach et al.~\cite{Hasselbach1988} that was especially optimized to be less sensitive to such dephasing mechanisms. The compact design is shielded by a surrounding magnetic mu-metal tube and all beam alignment is performed by electromagnetic deflection components to prevent any mechanical moving parts.\\
\\
The introduction of the SAT source is necessary for a sufficient coherent signal rate in the Aharonov-Bohm measurements. In former biprism interferometers for electrons or ions, other sources had been applied. They are compared in \cite{Schuetz2014}. In the ion interferometer of Hasselbach et al.~\cite{Hasselbach1998a,Hasselbach1996,Maier1997,Hasselbach2010a} so-called ''supertip'' sources \cite{Jousten1988} were used. It turned out that these tips showed low detection rates when installed in such a setup. Signal acquisition times of about \unit[15]{min} had to be taken into account \cite{Maier1997}. The low detection rate was a major obstacle for further experiments in the field of Aharonov-Bohm physics. This problem may be solved by SATs. In separate measurements \cite{Kuo2008,Schuetz2014} it was determined that these tips provide a brightness that is at least one order of magnitude higher than the one of the supertip sources. Thereby ''brightness'' is defined as the emitted ion current normalized to the emitting tip area (with a SAT diameter of \unit[0.3]{nm} \cite{Kuo2008}), the emission angle and the applied ionization gas pressure.

\begin{figure}[t]
%\centerline{\scalebox{0.5}{\includegraphics{figure2.eps}}}
\centerline{\scalebox{1.0}{\includegraphics{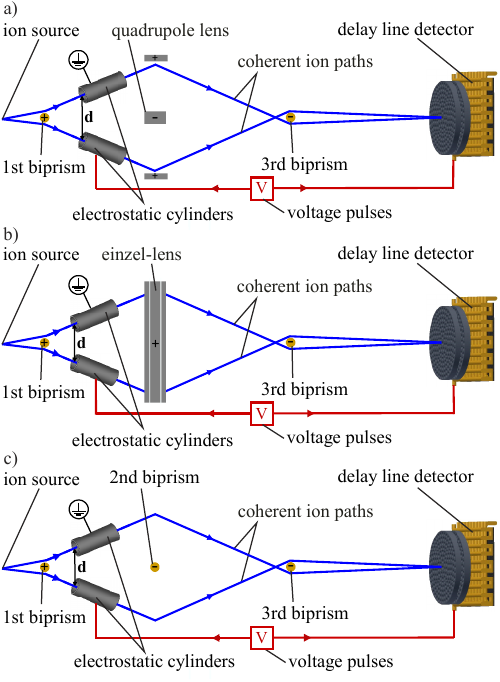}}}
\caption{(Color online) Proposed proof of the type I electric Aharonov-Bohm effect \cite{Aharonov1959} with a continuous ion beam and three possible schemes for the beam path separation (not to scale). The ion beam gets coherently divided by a positively charged first biprism. After a separation of $d=$~\unit[100]{\textmu m} they pass two metal cylinders. Then the beams get deflected by either a) a quadrupole lens (scheme: BP1-QP-BP3), b) an einzel-lens (scheme: BP1-EL-BP3) or c) the second biprism (scheme: BP1-BP2-BP3). A further negatively charged biprism combines and superposes the beam again to form an interference pattern after magnification (not shown) on a delay line detector. Applying a short voltage pulse on one of the cylinders, while the other one is grounded, shifts the phase of the interference pattern of the ion wave being inside the cylinders during the duration of the pulse. This signal can be selected due to the good time resolution of the delay line detector, which is correlated with the cylinder pulses.} 
\label{fig:elAB}
\end{figure}

\section{Concept for the measurement}
\label{elABeffect}

A full quantum mechanical description of the electric Aharonov-Bohm effect \cite{Aharonov1959} is given in \cite{Weder2011}. The beam originating from a SAT emitter gets coherently divided. The partial waves propagate through tiny metal cylinders and get combined on the detector where they interfere. In the original idea of Aharonov and Bohm \cite{Aharonov1959} an electron source is pulsed to emit wave packages smaller than the length of the metal cylinders. Inside the tubes, the particles are shielded against any fields from outside. As soon as the wave packages are inside, different potentials for the two tubes will be applied with the deviation $V$. They will be switched off before the electrons leave the cylinders. Thereby the electrons are exposed to scalar potentials, but not to electric fields. This causes a different phase shift of the partial waves, which can be observed in a shift of the interference pattern at the detector by~\cite{Aharonov1959}

\begin{equation}
\Delta \Phi_{el} = \frac{e}{\hbar} \int V dt~,
\label{eq:elAB}
\end{equation}

where it is integrated over the voltage pulse duration time $t$. One of the reasons why this original proposal of Aharonov and Bohm could not be verified up to now, is the high velocity of the electrons in the interferometer. Under normal conditions in a biprism interferometer experiment, the electrons get emitted by the metal tip with energies between 0.5 and \unit[3]{keV}.  
Lower energies could in principle be realized by slowing down the electrons with a counter electrode. But due to charging effects and electromagnetic noise it gets more and more demanding to control the beam and to maintain coherence. With a typical emission energy of \unit[1]{keV}, electrons have a velocity of \unit[$\sim 2\times10^7$]{m/s}. To apply realistic experimental conditions, the length of the tubes inserted into the beam paths is set to be \unit[3]{mm}. A comparable tube with such a length was successfully implemented into one beam path of an electron interferometer by Schmid et al. \cite{Schmid1985}. The electrons spend only a time of \unit[$\sim$~150]{ps} in such a cylinder. This is rather short to apply a full voltage pulse on one of the cylinders. Although it is not impossible since \unit[100]{ps} pulsers are commercially available. However, as we will outline below, it is additionally necessary to assure that only those electrons are selectively counted in the detector which have been in the cylinder when the pulse was on. This is to our knowledge technically to date not feasible within an accuracy of \unit[100]{ps}. \\
\\
With ions the situation is more comfortable. Emitting a beam of hydrogen dimers H$_2^+$ from a SAT at a voltage of \unit[3.8]{kV} \cite{Kuo2008} corresponds to an ion velocity of only \unit[$\sim 6 \times 10^5$]{m/s}, due to the larger mass comparing to electrons. The hydrogen ions would therefore spend a time of \unit[5]{ns} in a \unit[3]{mm} long cylinder. Voltage pulses with widths around \unit[1]{ns} are feasible to create with modern pulse generators. Applying electrons with such a low velocity instead of ions would require to slow them down to \unit[$\sim$ 1]{eV}, which is to date technically not feasible in a biprism interferometer setup. Therefore electrons are not suitable to prove the electric Aharonov-Bohm effect in such a device.  \\

\begin{figure*}
\centering

    \begin{minipage}{0.5\textwidth}
    \includegraphics[width=\textwidth]{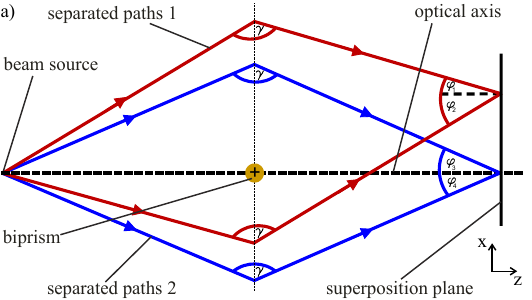}
    \end{minipage}
    %\hfill
    \hspace*{0.5cm}
    \begin{minipage}{0.3\textwidth}
    \includegraphics[width=\textwidth]{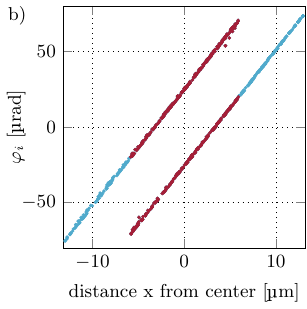}
    \end{minipage}

\caption{(Color online) a) Sketch of electron beam paths separated and deflected by an electrostatic biprism that overlap in a superposition plane. Each beam path gets deflected by the same angle $\gamma$ independent of its distance normal to the biprism wire. For that reason the superposition angle $\varphi$ of two paths in a certain plane is constant for all paths, determining the interference pattern periodicity $s$ by the relation $s=\lambda_{dB} / \varphi$. In a classical computer simulation the relative angles $\varphi_i$ of all particle pathways towards the perpendicular of the $xy$-plane at the entrance of the first magnifying quadrupole lens can be determined. In this sketch two of them are schematically illustrated with the angles $\varphi_1$ and $\varphi_2$ or $\varphi_3$ and $\varphi_4$. At any position in the superposition plane the sum of their values is constant: $\varphi_1 - \varphi_2 = \varphi_3 - \varphi_4 = \varphi$. Here it needs to be considered, if $\varphi_1$ (or $\varphi_3$) has a positive algebraic sign than $\varphi_2$ ($\varphi_4$) has a negative one. b) Simulation data for an electron interferometer experiment \cite{Schuetz2014} plotted according to the method described in the text. The data shows the crossing of the simulated electron trajectories in the plane at the entrance of the quadrupole lens. Each line corresponds to a separated beam path. The amount of superposition between the partial beams is printed in red.}
\label{Elektronen}
\end{figure*} 

As mentioned, Aharonov and Bohm suggested a pulsed coherent electron source \cite{Aharonov1959}. This is in principle possible by irradiating a pulsed femtosecond laser on a field emission tip set on a voltage. Such a source setup was realized in \cite{Hommelhoff2006} and emits femtosecond pulses of free electrons.
It is questionable, if this scheme can be transferred to ion emission. Therefore, it is experimentally more convenient to keep the ion emission continuous and perform a temporal selection of the ion events in the detector. The proposed scheme is illustrated in fig.~\ref{fig:elAB}. A pulse from a pulse generator will be separated into two correlated signals. One of them is attenuated and used to pulse one of the Aharonov-Bohm tubes. The other part of the separated pulse triggers a logic module in the delay line detector. It is able to correlate detection events with a certain delay to the trigger signal. The delay is set such that only those ions are counted that have been inside the metal tube while the electric potential on the tube was switched on. The time resolution of the detector needs to be in the same order as the pulse duration. This is feasible with a modern delay line detector \cite{Jagutzki2002} where a temporal resolution of about one nanosecond can be achieved.\\
\\
By selecting the ions on the detector which have been in the cylinders during the voltage pulse, only a tiny fraction of the ions from the source will be counted. The question arises if there is enough signal left to observe the Aharonov-Bohm phase shift within a reasonable integration time of several minutes. During significantly longer signal acquisition dephasing mechanisms such as temperature drifts, mechanical or electromagnetic noise may shift the phase of the pattern. The count rate on the detector depends significantly on the efficiency of the beam path separation scheme. We therefore performed computer simulations of the classical ion beam paths in the interferometer in the next section and compared the three different beam separation schemes shown in fig.~\ref{fig:elAB}. Thereby, the superposition angle and the fraction of the coherent ions counted on the detector relative to the emission from the SAT could be determined. This leads to the expected interference fringe period, the required pattern magnification and, in combination with performance data of SATs \cite{Kuo2008}, the total detection signal.

\section{Simulations of the ion beam paths}
\label{simulation}

To limit the signal acquisition time for the measurement of the electric Aharonov-Bohm effect the optimal setup needs to be found where a maximal fraction of the ions from the coherent source reaches the detector. Thereby, the beam separation and the distances between the components are important because they determine the superposition angle. A smaller distance e.g.~between the superposing last biprism and the entrance of the magnifying quadrupoles results in a larger minimal superposition angle and a smaller fringe period. This creates a necessity for a larger magnification to resolve the pattern on the detector which causes a loss of signal. It is therefore required to simulate the performance of the setup in different configurations. Until now a variety of beam path separation schemes have been realized for electron interferometers in the literature \cite{Sonnentag2007a,Mollenstedt1962,Schmid1985,Hasselbach1993b}. We choose to simulate and compare the three different schemes shown in fig.~\ref{fig:elAB} a-c) with the program {\it Simion} \cite{Simion} for $3\times 10^6$ H$_2^+$ ions and an emission energy of \unit[3.8]{keV}.\\

This program is able to calculate the classical trajectories of the particles. However, we can draw conclusions about their quantum behaviour by determining the superposition angle $\varphi$ of the combined beam at the entrance of the first magnifying quadrupole. This is justified if the coherent illumination is larger than the width (in $x$-direction) of the overlapping area of the two separated partial beams (which is in the range of \unit[1-2]{\textmu m} for H$_2^+$ ions at \unit[3.8]{keV} before magnification). The coherently illuminated area is determined by the angular coherence relation $\alpha \le \frac{\lambda_{dB}}{2 \, \epsilon}$, where $\alpha$ is the opening angle of the coherent emission, $\lambda_{dB}$ the de Broglie wavelength of the particle and $\epsilon$ the source size. The relation defines the maximal emission angle in which the ions are still coherent. Typical values for the proposed experiment are $\epsilon\sim$~\unit[0.3]{nm} \cite{Kuo2008}, $\lambda_{dB}=$~\unit[0.33]{pm} and $\alpha=$ \unit[$5.5\times10^{-4}$]{rad}. The coherently illuminated area in case that all components are grounded has a diameter of $\sim$~\unit[11]{\textmu m} at the first biprism and $\sim$~\unit[131]{\textmu m} at the entrance of the quadrupole. \\
\\
The interference pattern period $s$ can be calculated with the relation: $s=\lambda_{dB} / \varphi$. The matter wavelength $\lambda_{dB}$ is determined by the relation: $\lambda_{dB} = h / \sqrt{2 m e U}$, with the electron or ion mass $m$ and the emission voltage of the source $U$. Our method to extract the beam path superposition angle $\varphi$ from our simulations is schematically illustrated in fig.~\ref{Elektronen} a). It is a fundamental feature of a biprism that it deflects all electrons/ions in the near field by the same angle $\gamma$ independent on the distance normal to the biprism fiber \cite{Mollenstedt1956a}. Therefore, all beam paths that overlap in the $xy$-plane after a certain distance $z$ include the same superposition angle $\varphi$ (see fig.~\ref{Elektronen} a)). However, in our simulation we calculated the angles $\varphi_i$ between an arbitrary beam path and the normal perpendicular to the $xy$-plane at the $z$-position directly before the first magnifying quadrupole (superposition plane in fig.~\ref{Elektronen} a)). In fig.~\ref{Elektronen} b) all angles $\varphi_i$ are plotted versus the distance $x$ from the center in the $x$-direction. All data points gather along two parallel straight lines. The two lines yield the two separated partial beams that combine. The overlapping of the lines in the $x$-direction indicates the superposition of the two beams. Subtracting the two values $\varphi_i - \varphi_j$ at any position $x$ yields the superposition angle $\varphi$ which is constant (within the error of our simulation) for any point in the overlapping region. Therefore, we extracted the superposition angle $\varphi$ along this region and averaged the data to determine the error in our simulation.\\

\begin{figure}[tb]
%\centerline{\scalebox{0.5}{\includegraphics{figure4.eps}}}
\centerline{\scalebox{1.0}{\includegraphics{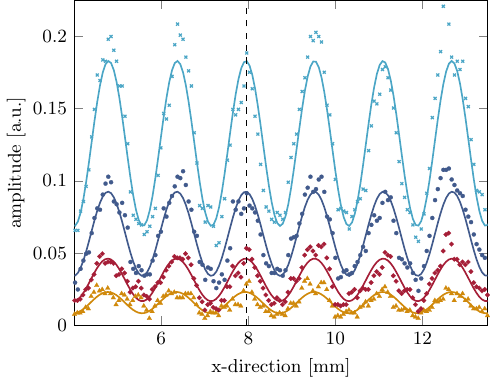}}}
\caption{(Color online) Averaged fringe patterns recorded in an electron biprism interferometer \cite{Schuetz2014,Hasselbach1998a,Hasselbach1996,Maier1997,Hasselbach2010a} with the total signal acquisition on the whole detection area of $2 \times 10^3$ (yellow triangles), \mbox{$5 \times 10^3$} (red squares), $1 \times 10^4$(dark blue dots) and $2 \times 10^4$ (light blue crosses) counts on a circular MCP-detector with a diameter of \unit[40]{mm}. The amplitude was determined by integration of the single electron events along the fringe direction in the recorded interferogram and divided by the number of pixel-columns on the detector. The data shown is a rectangular section and averaged over three adjacent data points. It is fitted by a sinusoidal fit that demonstrates the possibility to determine the correct phase for a total signal on the detector as low as $2 \times 10^3$ counts.} 
\label{fig:electroncounts}
\end{figure}

Before we apply this method on the proposed Aharonov-Bohm experiment, we demonstrate that it provides correct pattern periods by a comparison between an experimentally determined interferogram for electrons and our simulation. Thereby, we simulated the electron trajectories in the interferometer setup described in \cite{Schuetz2014}. A point source emits  $3 \times 10^6$ electrons with a fixed energy of \unit[1.55]{keV}. The de Broglie wavelength of the electrons can therefore be calculated to be \unit[$\lambda_{dB} = 3.1 \times 10^{-11}$]{m}. The emission angle is Gaussian distributed with a full width at half maximum (FWHM) solid angle of 6.15$^\circ$. The biprism has a diameter of \unit[400]{nm} and is set on a voltage of \unit[0.5]{V}. Only the main components are considered in the simulation which are the biprism and the quadrupole. All other parts in the experiment were neglected, such as the deflection electrodes, the Wien filter or the image rotator. A perfect alignment of the beam to the center is considered. The angles $\varphi_i$ as a function of the $x$-coordinate directly before quadrupole magnification are determined as described above and plotted in fig.~\ref{Elektronen} b). It can be deduced that the beam was superposed by \unit[11.5]{\textmu m} with an average superposition angle of $\varphi=$ \unit[50.44($\,\pm \, 0.9$)]{\textmu rad}. This corresponds to an interference stripe distance of \unit[$s = 615$($\,\pm \, 11$)]{nm}. The quadrupole magnification was M $=2828$ leading to a simulated pattern periodicity on the detector of \unit[1.74($\,\pm \, 0.03$)]{mm}. In fig.~\ref{fig:electroncounts} the experimentally determined data \cite{Schuetz2014} is shown for $2 \times 10^3$, $5 \times 10^3$, $1 \times 10^4$ and $2 \times 10^4$ counts on the detector. A sinusoidal fit reveals a period of \unit[1.58($\,\pm \, 0.01$)]{mm} that agrees reasonably well with our simulation. The deviation towards the measurement can be assigned to the alignment of the beam by the deflection electrodes, the grounded apertures at the entrance of all elements in the beam path and the uncertainty in the distances between the components in the experiment. Due to complexity and computational effort, these issues have been neglected in the simulation. After this test our method was applied to calculate the expected performance of the more complex Aharonov-Bohm setup.\\

\begin{figure*}[t]
%\centerline{\scalebox{0.5}{\includegraphics{figure5.eps}}}
\centerline{\scalebox{1.0}{\includegraphics{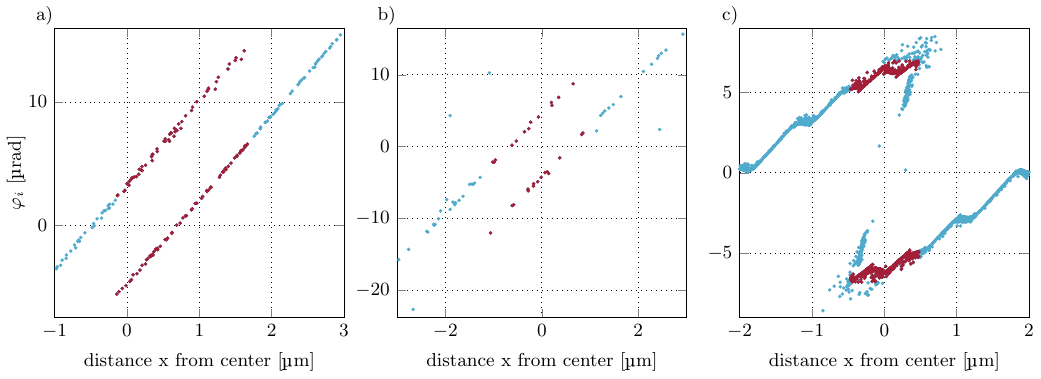}}}
\caption{(Color online) Results of the beam path simulations for the three different schemes to separate the beams by \unit[100]{\textmu m}. They are plotted according to the method explained in fig.~\ref{Elektronen} to extract the superposition angle at the entrance of the first magnifying quadrupole lens before magnification. The data in a), b) and c) correspond to the beam path separation illustrated in fig.~\ref{fig:elAB} a) by a quadrupole, b) an einzel-lens or c) the 2nd~biprism, respectively. } 
\label{fig:angles}
\end{figure*}

\begin{table*}
  \caption{Distances (in mm) along the beam path from the ion source to the center of the components that are included in the simulation for the three different beam path separation schemes. If a potential (in volts) other than zero was applied on the components, its value is written in the brackets. The voltage applied on the two opposing electrodes of the quadrupoles normal to the biprism fiber and parallel to it, respectively, are separated by a slash. }
  \label{tab:dimensions}
  \begin{tabular}{c|c|c|c|c|c|c|c|c|c}
    \hline
   \rule{0pt}{2.5ex} separation& 1st && 1st &  & 2nd & 3rd & 2nd & 3rd & \\
	 scheme  & biprism &  AB-tubes & quadrupole & einzel-lens & biprism & biprism & quadrupole & quadrupole &  detector \\ \hline
     \rule{0pt}{2.5ex}BP1-QP-BP3 & 20 & 60.5 & 87 &&& 117 & 282.25 & 307.55 & 592\\
     & (30) &&(352.3/-40)&&& (-69.57) & (-800/800) & (-2570/2570) &\\ \hline
     \rule{0pt}{2.5ex}BP1-EL-BP3  & 20 & 60.5 &  & 68.43  &  & 92    & 257.25   & 282.55 & 588 \\
     			&(30)&      &  &(1830)  &  &(-54.6)&(-900/900)&(-2040/2040)&  \\ \hline
     
     \rule{0pt}{2.5ex}BP1-BP2-BP3  & 20 & 60.5 & &  & 68     & 92     & 257.25   & 282.55 & 588 \\
     			 &(30)&      &&  &(-110.3)&(-80.76)&(-770/770)&(-3060/3060)&\\ \hline
     			
  \end{tabular}
\end{table*}

We simulated the interferometer in fig.~\ref{fig:biprism} b) and considered H$_2^+$ ions emitted by a point source with an energy of \unit[3.8]{keV}. Such as it was observed experimentally for a SAT source \cite{Schuetz2014,Kuo2008}, a Gaussian distributed emission into a FWHM solid angle of 1.5$^\circ$ was assumed. Again, the deflection electrodes, the Wien filter and the image rotator were neglected. These elements are only used to compensate for small misalignments in a real experiment and are not supposed to influence the calculated output signal. Therefore, the simulation includes the first biprism, the phase shifting tubes, which are both grounded, the focusing component (quadrupole lens or einzel-lens or 2nd~biprism), the 3rd~biprism and two quadrupole lenses. The distances between the elements are provided in table~\ref{tab:dimensions}. For each beam path separation scheme shown in fig.~\ref{fig:elAB} two simulations with $3\times 10^6$ ions were performed. In the first simulation we plotted the angles $\varphi_i$ at the entrance of the magnifying quadrupole lens versus the distance to the beam center in $x$-direction according to the method described above. The data is presented in fig.~\ref{fig:angles} for a) a quadrupole lens as the focusing element, b) an einzel-lens and c) the second biprism. From this data we calculated the results exhibited in table~\ref{tab:results}. The superposition angles are in the range of \unit[10]{\textmu rad} resulting in interference pattern periodicities of several ten nanometers. We choose the magnification such, that the periodicity on the MCP-detector is similar to the experimental situation in fig.~\ref{fig:electroncounts}. The H$_2^+$ ion interference is therefore expected to provide a comparable interferogram. In a second simulation we determined the total ions hitting the detector with a diameter of \unit[40]{mm}. This number reveals the fraction of the originally emitted $3 \times 10^6$ ions from the source that forms the pattern on the detector. Multiplying it with the measured emission rate of \unit[35]{pA} from the iridium SAT at \unit[3.8]{kV} by Kuo et al.~\cite{Kuo2008} gives the expected H$_2^+$ count rate in our proposed interferometer.\\
 
\begin{table*}
  \caption{Results of the beam path simulations for the three different separation schemes shown in fig.~\ref{fig:elAB}. The values are calculated from the data in fig.~\ref{fig:angles}. The amount of the superposition between the partial beams, the superposition angle $\varphi$ and the interference pattern periodicity are determined at the entrance of the first magnifying quadrupole lens. The expected H$_2^+$ count rate is calculated by combining the simulation for the amount of ions on the detector with the measured SAT emission rate from Kuo et al.~\cite{Kuo2008} at a hydrogen ionization gas pressure of \unit[$10^{-4}$]{Torr}.}
  \label{tab:results}
  \begin{tabular}{c|c|c|c|c|c|c|c}
\hline
     \rule{0pt}{2.5ex} separation & beam super- & superposition & periodicity before &  periodicity on & total signal on & H$_2^+$ count & AB signal \\
	scheme & position [\textmu m] & angle [\textmu rad] & magnification [nm] & detector [mm] & detector [ions] & rate [ions/s] & acquisition [s] \\ \hline 
   \rule{0pt}{2.5ex} BP1-QP-BP3 & 1.76 & 7.98($\,\pm \, 0.22$) & 41.0($\,\pm \, 1.1$) & 1.32($\,\pm \, 0.04$) & 58 & 4215 & 540 \\
    BP1-EL-BP3 & 1.71 & 8.80($\,\pm \, 0.58$) & 37.2($\,\pm \, 2.4$) & 1.04($\,\pm \, 0.07$) & 22 & 1599 & 1414 \\
    BP1-BP2-BP3 & 0.92 & 11.97($\,\pm \, 0.35$) & 27.3($\,\pm \, 0.8$) & 1.21($\,\pm \, 0.04$) & 395 & 28703 & 80 \\
    \hline
  \end{tabular}
\end{table*}

\begin{figure*}[t]
%\centerline{\scalebox{0.5}{\includegraphics{figure6.eps}}}
\centerline{\scalebox{1.0}{\includegraphics{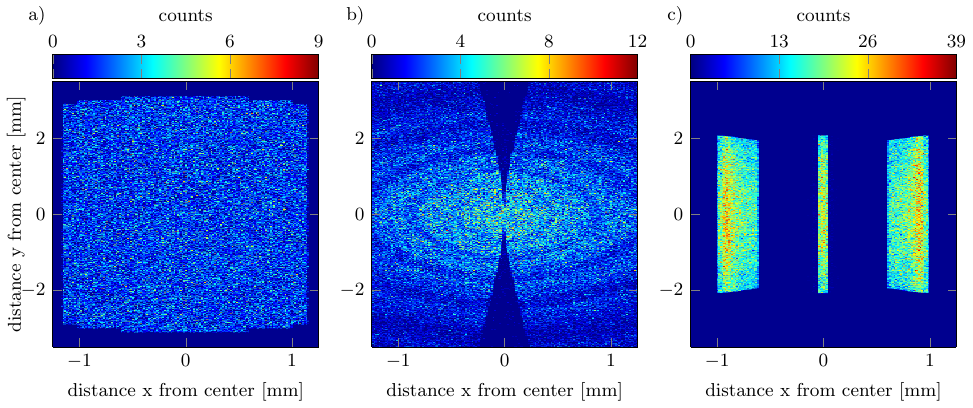}}}
\caption{Simulated intersection of the ion pathways with the $xy$-plane perpendicular to the beam axis at the entrance of the magnifying quadrupole lenses. a), b) and c) describe the situation for a quadrupole, an einzel-lens or with the 2nd~biprism as the focusing element in the beam path, respectively. The beam is preselected by a \unit[1]{mm} aperture before the phase shifting tubes. Since the third biprism focusses the beam only in the $x$-direction, the spreading of the ions in the $y$-direction by the einzel-lens cannot be compensated leading to a significant loss of signal in the direction parallel to the biprism compared to the other two separation schemes. The circular ring structures in b) are believe to be computational artifacts due to the limited resolution in our simulation. They are not relevant for our calculations, since we only consider ions in the small coherently illuminated area close to the center where no ring structures are observed.} 
\label{fig:beforeQP}
\end{figure*}

The dimensions of the biprism fiber ($\sim$ \unit[400]{nm}) is much smaller than the distance between the two grounded plates (\unit[4]{mm}) where it is positioned in the center. To maintain a reasonable computation time, it was necessary to limit the resolution of the fiber geometry for the calculation of the potential between the fiber and the plates. However, this influenced especially the simulation in the scheme with three biprisms in fig.~\ref{fig:angles} c) and caused a stepped form of the data. Tests revealed that the points in the field between the two stepped lines are misguided ions that followed a pathway in close vicinity of the fiber surface and are therefore computational artifacts from the limited resolution. With perfect round fiber symmetry they would be absorbed on the biprism surface. We excluded such points and used only the ion incidents marked in red in fig.~\ref{fig:angles} c) within the superposition region for the analysis of the data. Their fraction is about two thirds of the ions that reach the detector. Therefore, the total signal on the detector in the scheme with three biprisms and the expected H$_2^+$ count rate in table~\ref{tab:results} is corrected by a factor 0.66.  \\
The comparison between the three beam separation schemes in table~\ref{tab:results} clearly reveals that the expected count rate is highest with the 2nd~biprism as the focusing element. The reason for such a behaviour can be found by plotting the ion pathways intersections in the $xy$-plane perpendicular to the beam axis just before quadrupole magnification. This is done in fig.~\ref{fig:beforeQP}. Due to the round symmetry of the einzel-lens the originally straight shadow of the biprism fiber is distorted. Therefore, only the ions in the center close to the beam axis overlap and the spreading of the ions is larger. For the quadrupole lens the ions get also spread over a wider area compared to the situation with the second biprism as the focusing element. We therefore use this scheme and its count rates for the further calculation concerning the measurement of the electric Aharonov-Bohm effect.\\

\section{Expected signal acquisition time}
\label{count rate}

\begin{figure}[t]
%\centerline{\scalebox{0.5}{\includegraphics{figure7.eps}}}
\centerline{\scalebox{1.0}{\includegraphics{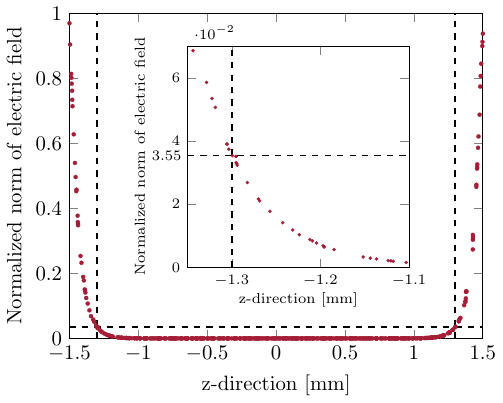}}}
\caption{Simulation of the norm of the electric field in the proposed Aharonov-Bohm tubes along the two separated ion pathways \unit[50]{\textmu m} left and \unit[50]{\textmu m} right of the optical axis in the $xz$-plane. The field strength is normalized to the highest field value near the entrance of the tubes. In the simulation only one tube is set on a voltage, whereas the other one is grounded, such as described in the text. It can be determined that only 3.55\% of the field is left after a distance of \unit[200]{\textmu m} from both edges of the \unit[3]{mm} long tube at \unit[$z = \pm 1.3$]{mm}. Only the tube length within this shielded region (\unit[2.6]{mm}) is used in the calculations for the Aharonov-Bohm signal to guarantee an almost field free type I Aharonov-Bohm measurement. Inset: Magnification of the simulated field close to \unit[$z = -1.3$]{mm}}
\label{fig:Comsol}
\end{figure}

Our simulations demonstrate that with the bright single atom tip (SAT) source \cite{Fu2001,Kuo2006a,Kuo2008} a reasonable high coherent H$_2^+$ signal rate will be detected in the proposed interferometer even after a beam separation and the implementation of two Aharonov-Bohm tubes. However, the expected rates in table~\ref{tab:results} will be decreased significantly in the pulsed mode proposed in fig.~\ref{fig:elAB} for the measurement of the electric Aharonov-Bohm effect. In this section we determine if there is still enough signal left to observe the phase shift after a reasonable acquisition time. \\
\\
To be a proof of the type I electric Aharonov-Bohm effect, it is essential to assure, that the counted particles do not encounter any fringe fields at the beginning and the end of the tubes. To determine the expected shielding properties of the proposed Aharonov-Bohm tubes, we performed a simulation of the electric field around and in the tubes with the program {\it Comsol} \cite{Comsol}. Thereby an electrostatic potential is applied on one of the tubes, whereas the other one is grounded. The resulting normalized norm of the electric field along the two \unit[100]{\textmu m} separated ion pathways left and right of the optical axis in the $xz$-plane is shown in fig.~\ref{fig:Comsol}. It can be observed that the penetrating norm of the electric field at a distance of \unit[200]{\textmu m} from the tube edges inside the tubes is decreased down to 3.55\% compared to the field near the entrance of the tubes. For that reason we assume in our calculations of the expected Aharonov-Bohm signal a shielded length of \unit[2.6]{mm} within the \unit[3]{mm} long tubes.\\
However, the proposed experiment is not performed in an electrostatic mode. Our scheme operates with short voltage pulses with fast rise and fall times. The following argument can point out that the tube can still provide the same shielding as in the electrostatic case. The stray field with an applied pulse rise time of \unit[100]{ps}, as considered in this proposal, can be compared to the field inside the tubes under the influence of a \unit[5]{GHz} external electromagnetic field. In such a wave field the time between the maximum and the minimum of the electromagnetic oscillation (half of the wavelength) is \unit[100]{ps}. The wavelength of the \unit[5]{GHz} radiation is about \unit[60]{mm}. This is much longer than the maximal dimension of the tubes which is \unit[3]{mm}. For that reason the tubes with the ion inside are exposed to a quasi-stationary electromagnetic field \cite{Wolfsperger2008}. The shielding properties are comparable to the stationary, electrostatic case because the field changes on a significant larger length scale than the dimensions of the tube. The situation would be different if the wavelength of the field gets smaller than the tube dimensions (pulse length below \unit[10]{ps}). Then there is still shielding, but wave reflections, absorptions and transmissions need to be considered \cite{Wolfsperger2008}.\\

To estimate the interference signal expected in the pulsed mode, we start with the hydrogen count rate of 28703 ions per second as determined in the last section. With a total distance between the source and the detector of \unit[588]{mm} and a velocity of \unit[$6 \times 10^5$]{m/s}, a H$_2^+$-ion spends \unit[$\sim$ 1]{\textmu s} inside the beam path. If we pulse the cylinder with \unit[1]{MHz}, we assure that every ion encounters only one pulse during that time. When this pulse is applied, only in a few cases the ion is inside the cylinder and therefore recorded by the detector. This fraction is given by the shielded cylinder length (\unit[2.6]{mm}) divided by the length of the interferometer (\unit[588]{mm}). Therefore $\sim$~127 ions per second are inside the tubes when a pulse is applied and are counted by the detector. Considering the high emission rate (\unit[35]{pA}) of the SAT, it could be objected to be unrealistic to distinguish such a low count rate from the noise level. However, the counting on the detector is correlated with the voltage pulse on the tubes. For that reason the correlation time window in which ions are counted on the detector can be around \unit[1]{ns}, excluding most of the noise. \\
\\
The expected Aharonov-Bohm phase shift depends, according to eq.~\ref{eq:elAB}, on the voltage difference between the two cylinders and on the pulse width. Assuming a width of \unit[400]{ps}, which can be produced by commercially available pulse generators, yields a phase shift of $2\pi$ if a voltage difference of \mbox{\unit[$\sim$ 10]{\textmu V}} is applied. For a convincing observation of the full electrostatic Aharonov-Bohm phase shift of $2\pi$ we assume to record five interference patterns in steps of \unit[2]{\textmu V} applied on the cylinders.\\
Such a control of the pulses on the \textmu V level is not trivial. Great care needs to be taken to shield the whole interferometer from electromagnetic noise. However, this is anyway a requirement in ion or electron interferometry to avoid dephasing of the interference pattern. We propose to use a pulse generator creating pulses with a height of \unit[10]{V} and rise times in the \unit[100]{ps} range. It can be connected to e.g.~two commercially available fixed precision attenuators with \unit[35]{dB} each for frequencies around \unit[10]{GHz}. The resulting \unit[70]{dB} attenuation will result in \unit[1]{\textmu V} pulses that are applied to the Aharonov-Bohm tubes. Since the attenuators are passive elements, they are not supposed to introduce noise or influence the length of the pulses. To avoid reflections it is important to connect the end of the contacted Aharonov-Bohm tube with ground by a \unit[50]{$\Omega$} termination impedance.\\
\\
A further question also needs to be discussed concerning the duration of the whole proposed Aharonov-Bohm measurement and if the apparatus is stable enough during that time. In \cite{Hasselbach1998a,Hasselbach1996,Maier1997,Hasselbach2010a} the total acquisition time for the measured helium ion interferogram was 15 minutes. Therefore, no significant dephasing mechanism was observed on that time scale. 
We determined the minimal counts needed to extract the phase of a single interference pattern in the biprism electron interferometer mentioned in section~\ref{simulation} \cite{Schuetz2014}. Several interferograms with signal acquisition of $2 \times 10^3$, $5 \times 10^3$, $1 \times 10^4$ and $2 \times 10^4$ counts were analysed in fig.~\ref{fig:electroncounts}. The data represents the integrated and normalized signal along the interference fringes on the detector. Each pattern is fitted by a sine curve. All patterns are in phase, demonstrating that only $2 \times 10^3$ coherent particles on the detection area are sufficient to clearly determine the phase of an interferogram.
Comparing this outcome to the expected hydrogen ion count rate of $\sim$~127 ions per second, the integration time for one pattern at a specific cylinder voltage would be about \unit[16]{s}. Recording five such interferograms to cover the full Aharonov-Bohm phase difference of $2\pi$ would therefore take approximately \unit[80]{s}. Comparing this time with the mentioned 15 minutes stability measured by \cite{Hasselbach1998a,Hasselbach1996,Maier1997,Hasselbach2010a} in a helium ion interferometer, we conclude, that dephasing effects will presumably not influence the outcome of our proposed measurement. As presented in table~\ref{tab:results}, the acquisition time for the separation scheme with a quadruople lens can be determined the same way and is expected to be \unit[540]{s} and for the einzel-lens it is \unit[1414]{s}. \\
\\
In a real experimental situation it could be difficult to perfectly align all three biprisms parallel. In that case it is possibly beneficial to replace the second biprism by an einzel-lens since its performance is rotationally symmetric. Even though the simulated signal rate on the detector is lower by a factor of $\sim$ 18. It is therefore a matter of the experimentalist to judge if alignment or stability issues are of greater concern.

\section{Conclusions}

When Aharonov and Bohm suggested their famous two experiments \cite{Aharonov1959}, a new door was opened to fundamental research in quantum physics. They predicted a direct physical impact of an electric scalar or magnetic vector potential in absence of any fields. Various experiments followed \cite{Chambers1960,Mollenstedt1962,Tonomura1986,Matteucci1985,Cimmino1989,Allman1992,Lee1998,Batelaan2009}, but a direct proof of the type I electric Aharonov-Bohm effect was not possible yet due to technical limitations. \\
In this paper, we propose such an experiment in a biprism interferometer with hydrogen ions. The experimental setup is related to the first electron interferometer with a single atom tip beam source \cite{Schuetz2014,Kuo2006a,Kuo2008} and the first ion interferometer \cite{Hasselbach1998a,Hasselbach1996,Maier1997,Hasselbach2010a}. For the measurement of the Aharanov-Bohm effect a bright ion source and an efficient coherent beam path separation scheme are necessary. Therefore, we performed a computer simulation for a beam path separation of \unit[100]{\textmu m} in three different setups. We presented a method to calculate the period of the quantum mechanical interference pattern by determining the superposition angle between the  partial beams. The simulation method was tested with good agreement on an experimentally determined interferogram in a biprism electron interferometer \cite{Schuetz2014}.\\
\\
The simulations for the setup to measure the electric Aharonov-Bohm effect with hydrogen ions indicate that a separation scheme including three biprisms in combination with a single atom tip source \cite{Kuo2008} yield the highest count rate at the detector. 
To observe the Aharonov-Bohm phase shift two metal tubes set on different electric potentials need to be positioned around the separated partial matter waves. We discussed a new concept to avoid the pulsed particle operation originally proposed by Aharonov and Bohm \cite{Aharonov1959}. Thereby, a delay line detector is applied with a high temporal resolution. It is able to selectively measure those ions in the continuous beam that are relevant to observe the Aharonov-Bohm phase shift. Combining our simulations with the measured H$_2^+$-ion emission current \cite{Kuo2008}, we predict a total signal acquisition time of about \unit[80]{s} which is short enough to prevent dephasing effects.  \\
We also point out that with a few modifications, the setup could be used to measure possible deviations \cite{Silverman1993} in the magnetic Aharonov-Bohm effect due to the internal structure of ions compared to electrons. We conclude our studies with the finding that a combination of ion matter waves \cite{Hasselbach1998a,Hasselbach1996,Maier1997,Hasselbach2010a}, single atom tip sources \cite{Kuo2008}, time resolving detectors \cite{Jagutzki2002} and a new beam path separation scheme is technically feasible and will allow for the first time the direct proof of the type I electric Aharonov-Bohm effect.

\section{Acknowledgements}

This work was supported by the Deutsche Forschungsgemeinschaft (DFG, German Research Foundation) through the Emmy Noether program \mbox{STI 615/1-1.} A.R. acknowledges support from the Evangelisches Studienwerk e.V. Villigst.

\end{document}